\documentclass[10pt,preprint]{aastex}
\usepackage{amsmath}

\newcommand{\elie}{{\mathbf E}}

\newcommand{\win}{w_{\rm in}}
\newcommand{\wout}{w_{\rm out}}
\newcommand{\ain}{a_{\rm in}}

\newcommand{\aout}{a_{\rm out}}

\newcommand{\elik}{{\mathbf K}}

\slugcomment{Accepted in Ap. J.}

\shorttitle{Numerical potential and field in discs}
\shortauthors{Hur\'e \& Pierens}

\begin{document}

\title{ACCURATE NUMERICAL POTENTIAL AND FIELD IN RAZOR-THIN AXI-SYMMETRIC DISCS}

\author{Jean-Marc Hur\'e}
\affil{LUTh/Observatoire de Paris-Meudon (UMR 8102 CNRS),\\ Place Jules
Janssen, 92195 Meudon Cedex, France}
\affil{Universit\'e Paris 7 Denis Diderot,\\ 2 Place Jussieu, 75251
Paris Cedex 05, France}
\email{jean-marc.hure@obspm.fr}

\and 

\author{Arnaud Pierens}
\affil{LUTh/Observatoire de Paris-Meudon (UMR 8102 CNRS),\\ Place Jules
Janssen, 92195 Meudon Cedex, France}

\email{arnaud.pierens@obspm.fr}

              
\begin{abstract}

We demonstrate the high accuracy of the
density splitting method to compute the gravitational potential
and field in the plane of razor-thin, axially symmetric discs, as
preliminarily outlined in Pierens \& Hur\'e (2004). Because residual kernels in Poisson integrals are not $C^\infty$-class functions, we use a dynamical space mapping in order to increase the efficiency of advanced quadrature schemes. In terms of accuracy, results are better by orders of magnitude than for the classical FFT-methods.
\end{abstract} 


\section{Introduction and motivations}

Discs are ubiquitous objects in the Universe and span different velocity/length scales: galactic (stellar) discs, Active Galactic Nuclei discs, circumstellar discs, binary and circum-binary discs, sub-nebulae. For many of them, self-gravity plays role in their structure and dynamics and so, the gravitational potential and associated accelerations are required at a certain level of disc modeling. Solving the Poisson equation in {\it extended, continuous media} like gas discs is however not trivial practically, and it has occupied astrophysicists for many decades. For time-dependent simulations, fast but low accuracy algorithms are generally preferred. For steady state analysis, accuracy is more critical than computing time, as it allows for instance to characterize the precise connection between various branches of solutions (e.g. Hachisu 1986, Ansorg, Kleinw\"achter \& Meinel 2003).

 Many numerical methods have been proposed, but a very few uses the integral formalism. In a previous paper (Pierens \& Hur\'e 2004; hereafter Paper I), we have outlined a method to avoid the singularity in the Poisson kernel so that the field in the plane of razor-thin, axially symmetric discs can easily be accurately computed from elliptic integrals by a single radial quadrature. The motivation of the present paper is twofold. First, Paper I just touches the problem at the theoretical level without giving numerical examples and discussing the implementation and possible performances of the method. Also, the potential was not considered. This is done here. Second, we have recently realized that the classical FFT-method (e.g. Binney \& Tremaine 1987) which is among the most widespread methods, have a much lower precision and lower order of convergence, in comparison. We show actually here that the density splitting method can exhibit a high convergence order, depending on the quadrature rule, and the precision can easily reach the machine precision with a few tens sources points, at least for smooth surface density profiles.

We outline the splitting method for the gravitational potential and radial field in Section 2. We then stress the non-derivability of Poisson kernels in Section 3, and propose a space mapping. In Section 4, we illustrate the possible performances of the method on a test-case (namely, a finite size disc with exponentially decreasing surface density profile), with three different quadrature rules. We compare the accuracy of the method with the classical FFT-method in Section 5. A few concluding remarks are found in the last section.

\bigskip
\begin{figure}[h]
\epsscale{0.45}
\plotone{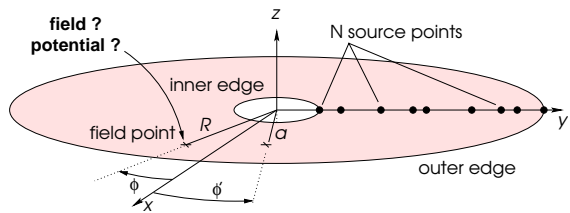}
\caption{Typical configuration and notations.}
\label{fig:config}
\end{figure}

\section{Outline of the density splitting method}

The gravitational potential $\Psi$ and radial component of the field $g_R = - \partial_R \Psi$ in the plane of a razor thin axially symmetric disc (see Fig. \ref{fig:config}) is\footnote{Matrix notation is employed only for compactness.} (e.g.
Durand 1964):
\begin{equation}
\begin{pmatrix}
\Psi\\
g_R
\end{pmatrix}
=\int_{\ain}^{\aout}\Sigma
\begin{pmatrix}
\kappa_\Psi\\
\kappa_g
\end{pmatrix}da,
\label{eq:psig}
\end{equation}
where  $\ain$ is the inner edge, $\aout$ is the outer edge, $\Sigma(a)$ is the surface density,
\begin{equation}
\begin{pmatrix}
\kappa_\Psi\\
\kappa_g
\end{pmatrix}
=-2G\sqrt\frac{a}{R}k\begin{pmatrix}
\elik(k)\\
\frac{1}{2R}\left[\elik(k) - \frac{\elie(k)}{\varpi}\right]
\end{pmatrix},
\label{eq:psig2}
\end{equation}
are the Poisson kernels, $\elik$ and $\elie$ denote the complete elliptic integral of the
first and second kinds respectively, $k=2\sqrt{aR}/(a+R) \le 1$ is their
modulus, and $\varpi=(a-R)/(a+R)$. As it is well known, these kernels diverges when $a \rightarrow R$, with the result that, in practice, neither $\Psi$ nor $g_R$ can properly be determined by direct integration for $R \in [\ain,\aout]$. As outlined\footnote{Only the field was considered in Paper I.} in Pierens \& Hur\'e (2004) (hereafter Paper I), such a singular behavior can be avoided if the surface density profile is split into two components according to
\begin{equation}
\Sigma(a)=\Sigma_0+\delta \Sigma(a,R),
\end{equation}
where $\Sigma_0 \equiv \Sigma(R)$ is the local value, and $\delta
\Sigma$ is the remainder (a function which depends on $a$ and $R$). Consequently, the potential and the radial field in the disc are given by
\begin{equation}
\begin{pmatrix}
\Psi\\
g_R
\end{pmatrix}
=
\begin{pmatrix}
\Psi^{\rm homo.}\\
g_R^{\rm homo.}
\end{pmatrix}
+
\begin{pmatrix}
\Psi^{\rm res.}\\
g_R^{\rm res.}
\end{pmatrix},
\end{equation}
where $\Psi^{\rm homo.}$ and $g_R^{\rm homo.}$ are analytical functions (proportional to  $\Sigma_0$; see Appendix A in Paper I for the field, and Appendix A in this paper for the potential). Terms $\Psi^{\rm res.}$ and $g_R^{\rm res.}$ correspond to the departure from the homogeneous disc. These are simply given by
\begin{equation}
\begin{pmatrix}
\Psi^{\rm res.}\\
g_R^{\rm res.}
\end{pmatrix}
=\int_{\ain}^{\aout}\delta \Sigma
\begin{pmatrix}
\kappa_\Psi\\
\kappa_g
\end{pmatrix}da.
\label{eq:psig}
\end{equation}

The point is that both $\delta \Sigma
\times \kappa_g$ and $\delta  \Sigma \times \kappa_\Psi$ are finite when
$a=R$, although $\kappa_\Psi \rightarrow \infty$ and
$\kappa_g \rightarrow \infty$ (see
Appendix B in Paper I for a proof), with
\begin{equation}
\begin{cases}
\left. \left(\delta \Sigma \times \kappa_\Psi \right) \right|_{a=R} = 0,\\
\\
\left. \left( \delta \Sigma \times \kappa_g \right) \right|_{a=R}= 2G \left(\frac{d\Sigma}{da}\right)_{a=R}.
\end{cases}
\label{eq:kernelsaisr}
\end{equation}
Note that, in this procedure, the radial derivative of the surface density  $\frac{d\Sigma}{da}$ is needed at each field point when computing $g_R^{\rm res.}$ from Eq.(\ref{eq:psig}). This is quite uncomfortable if the surface density is not defined analytically, but on a grid as in many simulations.

\begin{figure}[h]
\epsscale{0.4}
\plotone{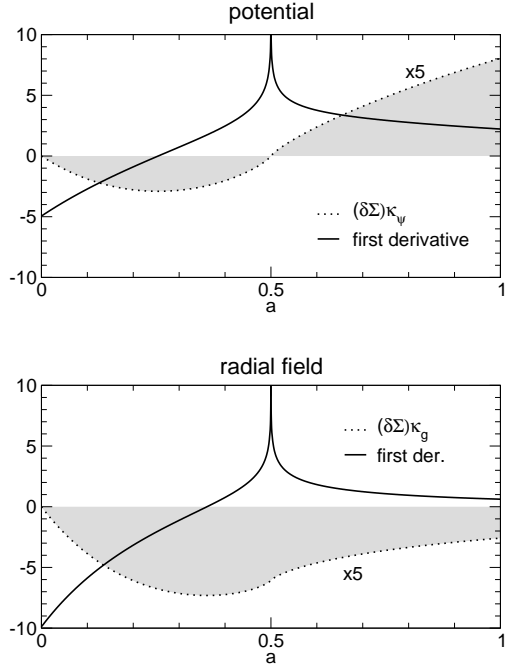}
\caption{{\it Top}: the function  $\delta \Sigma \times \kappa_\Psi$
versus $a$ ({\it dashed line}) and its first derivative ({\it plain line})
in a disc with inner edge $\ain=0$, outer edge $\aout=1$, and
exponentially decreasing surface density profile. The field point is
located at $R=0.5$. {\it Bottom}: same but for the radial field. Areas
under $\delta \Sigma \times \kappa_\Psi$ and  $\delta \Sigma \times
\kappa_g$ ({\it filled zones}) are the residual terms $\Psi^{\rm res.}$
and $g^{\rm res.}$ respectively.}
\label{fig:inftyder}
\end{figure}

\section{Residual integrands are not C$^\infty$-class functions}

Accuracy of residual terms is set by the scheme performing the numerical quadrature, provided the integrand is a well-behaved function. In the present problem, a difficulty arises because the residual integrands $\delta
\Sigma \times \kappa_g$ and $\delta  \Sigma \times \kappa_\Psi$ are continuous, but
not C$^\infty$-class functions (i.e. differentiable for all degrees of differentiation). Even, these are non-derivable (i.e. not C$^0$). This is easily understood from the first derivative
\begin{equation}
\frac{d}{da} \left( \delta \Sigma  \times  \kappa \right)  =
\frac{d \Sigma}{da} \times \kappa +  \delta  \Sigma \times \frac{d \kappa}{da},
\label{eq:dsk}
\end{equation}
where $\kappa$ denotes either $\kappa_\Psi$ or $\kappa_g$,
\begin{equation}
 \frac{d \kappa}{da} =  \frac{d \kappa}{dk} \times \frac{dk}{da}
\end{equation}
and
\begin{equation}
\frac{dk}{da} = \sqrt{\frac{R}{a}}\frac{(R-a)}{(a+R)^2} = - \frac{\varpi k}{2a}.
\end{equation}

We see that the first term in the right-hand-side of Eq.(\ref{eq:dsk}) brings a diverging contribution for $a=R$ since $d \Sigma /da$ can not be zero on the whole integration range. We thus have $\frac{d}{da} \left( \delta \Sigma  \times  \kappa \right)  \rightarrow \infty$ at the field point. This point is illustrated in Fig. \ref{fig:inftyder} which displays
the function $\delta \Sigma \times \kappa$ for both the potential and
field as well as their first derivative in a disc with exponentially
decreasing surface density profile. Although not visible at the scale of the graphs, there is a small ``knee'' just at the field point where the first derivatives are infinite. Further, because successive derivatives of
elliptic integrals {\it inevitably} produce the $\elik$-function which
is logarithmically diverging as $a \rightarrow R$, we conclude that
\begin{equation}
\left.\frac{d^n}{da^n}(\delta \Sigma \times \kappa) \right|_{a=R}
\rightarrow \infty \qquad \text{for any } n \ge 1.
\end{equation}

\begin{deluxetable}{lcc}
\tablecolumns{4} 
\tablewidth{0pc} 
\tablecaption{Values of the integrands $\delta \Sigma \times \kappa$ and its $a$-derivatives at the field point.} 
\tablehead{
\colhead{order of derivative} & \colhead{$\delta \Sigma \kappa_\Psi$ at $a=R$} &\colhead{$\delta \Sigma \kappa_g$ at $a=R$ }}
\startdata 
$0$th (function) & $0$ & $2G\left(\frac{\partial \Sigma}{\partial a}\right)_{a=R}$ \\
$1$st derivative & $\infty$ & $\infty$\\
$\ge 2$nd & $\infty$ & $\infty$ \\
\enddata 
\label{tab:deriva}
\end{deluxetable}

\begin{deluxetable}{ccc}
\tablecolumns{4} 
\tablewidth{0pc} 
\tablecaption{Same legend as for Tab. \ref{tab:deriva}, but when in the $w$-space.}
\tablehead{
\colhead{order of derivative} & \colhead{$\delta \Sigma \kappa_\Psi |w|$ at $w=0$} &\colhead{$\delta \Sigma \kappa_g |w|$ at $w=0$ }}
\startdata 
$0$th (function)& $0$ & $0$ \\
$1$st derivative & $0$ & $\mp 4G\left(\frac{\partial \Sigma}{\partial a}\right)_{a=R}$  \\
$2$nd derivative & $0$ & $\infty$ \\
$3$rd derivative & $\infty$ & $\infty$\\
$\ge 4$th&$\infty$ & $\infty$ \\
\enddata 
\label{tab:derivw}
\end{deluxetable}

Values of the integrands $\delta \Sigma \times \kappa$ and its $a$-derivatives at $a=R$ are summarized in Tab. \ref{tab:deriva}. It means that, if residuals terms are numerically determined following Eq.(\ref{eq:psig}) (i.e. by integration in the $a$-space), then most classical quadrature rules which are based on a certain fitting of the integrand by polynomials cannot behave in an optimal manner. This problem especially concerns high-order quadrature rules. We have no idea how to totally remove such a difficulty. It is possible however to reduce it partially by re-sampling the integrand in an appropriate way. For instance, if we consider the function $w(a)$ defined by
\begin{equation}
w(a)=\begin{cases}
- \sqrt{\frac{R-a}{\aout}} < 0 & \text{if} \quad a < R,\\
0 & \text{if} \quad R=a, \\
\sqrt{\frac{a-R}{\aout}} > 0 & \text{if} \quad a > R,\\
\end{cases}
\label{eq:w}
\end{equation}
then Eq.(\ref{eq:psig}) now reads
\begin{equation}
\begin{pmatrix}
\Psi^{\rm res.}\\
g_R^{\rm res.}
\end{pmatrix}
= 2\aout \int_{\win}^{\wout}\delta \Sigma
\begin{pmatrix}
\kappa_\Psi\\
\kappa_g
\end{pmatrix} |w|  dw,
\label{eq:psigdw}
\end{equation}
where $\win=w(\ain)$ and $\wout=w(\aout)$. The ``new integrands'' $\delta \Sigma \times \kappa_\Psi |w|$ and  $\delta \Sigma \times \kappa_g |w|$ are obviously regarded as a function of the new variable $w$. The advantage of this space mapping is twofold. First, it makes the first derivative finite; $w$-derivatives of the new integrands are summarized in Tab. \ref{tab:derivw} (see Appendix B for a detailed calculus). Figure \ref{fig:inftyderdw} displays the first, second and third derivatives, in the same conditions as for Fig. \ref{fig:inftyder}. Second, $\delta \Sigma \times \kappa_g |w|$ vanishes at the field point (since $w=0$). Hence, $d \Sigma/da$ is not more needed.

\begin{figure}[h]
\epsscale{0.4}
\plotone{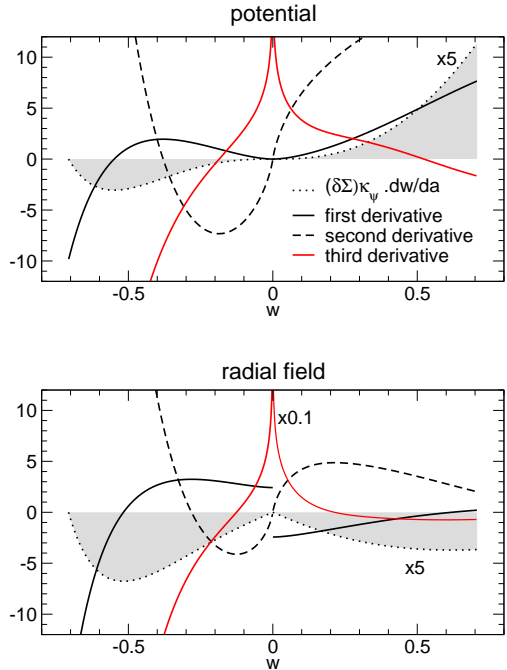}
\caption{Same legend as for Fig. \ref{fig:inftyder} but the integrands $\delta \Sigma \times \kappa_\Psi \times \frac{dw}{da}$ ({\it top})  and  $\delta \Sigma \times \kappa_g\times \frac{dw}{da}$ ({\it bottom}).}
\label{fig:inftyderdw}
\end{figure}

\section{Example of performance}

We briefly show the possible performances of the method through a typical example, namely a disc with $[\ain,\aout] = [0,1]$ and an exponentially decreasing surface density profile, as already considered above. Tests have concerned a large amount of disc models (various surface density profile, various axis ratios $\aout/\ain$). For simplicity, we shall discuss only the potential; results for the field are similar. We consider three different schemes to determine $\Psi^{\rm res.}$ at various radii $R$ inside the disc from Eq.(\ref{eq:psigdw}), namely:
\begin{itemize}
\item the composite trapezoidal rule (hereafter, the CT-rule), with $N$ source points. This is a 2nd-order accurate scheme (e.g. Press et al. 1992).
\item a $6$th-order, regular-spacing quadrature rule due to Gill \& Miller (1972), with $N$ source points (hereafter, the GM-rule). 
\item the Gauss-Chebycheff-Lobatto approximation (hereafter, the GCL-rule), with $N$ Chebycheff polynomials. This
collocation method would be the most efficient technique  (with a
spectral convergence) in the presence of a C$^\infty$-class integrand (e.g. Boyd 2001).
\end{itemize}

We measure the relative precision  $\Delta \Psi/\Psi$ on potential values $\Psi$ with the $\epsilon$-parameter defined as
\begin{equation}
\epsilon \equiv
\log_{10} \left|\frac{\Psi - \Psi^{\rm ref.}}{\Psi^{\rm ref.}} \right|,
\label{eq:errorindex}
\end{equation}
where $\Psi^{\rm ref.}$ is the reference value. Since the exact potential is not known for the case considered, reference values $\Psi^{\rm ref.}$ are obtained by considering a larger number of source points, as commonly done (e.g. Cohl \& Tohline 1999). We use double precision (DP) computer calculus so that the precision is limited to $\sim 2 \times 10^{-16}$ (that is $\epsilon \ge \epsilon_{\rm DP} \simeq -15.7$).

\begin{figure}[h]
\epsscale{0.4}
\plotone{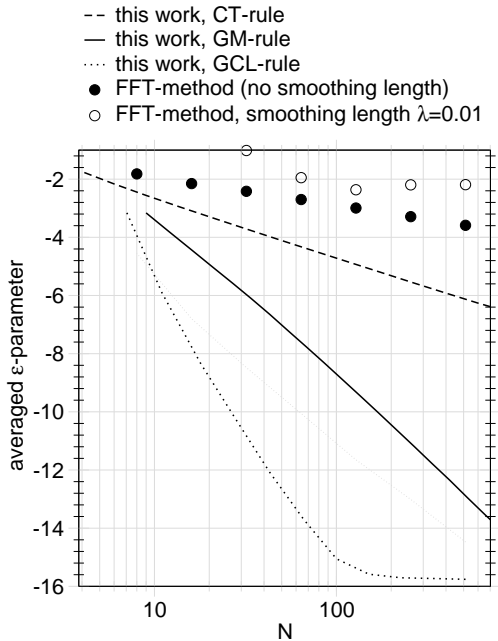}
\caption{Averaged $\epsilon$-parameter versus $N$ for the splitting methods and for the FFT-method with  smoothing length ({\it open circles}; $\lambda=0.01$) and without  smoothing length ({\it filled circles}). For the FFT-method, the disc inner edge is set to $10^{-2}$. See Fig. \ref{fig:errorsFFT} for the $\epsilon(R)$ and $N=128$.}
\label{fig:errors}
\end{figure}

Figure \ref{fig:errors} gives the results of the splitting method, namely $\langle \epsilon \rangle$ versus $N$ for the three quadrature rules listed above, where $\langle \epsilon \rangle$ is an averaged value, obtained by computing $\Psi$ at $2^7$ equally spaced positions spanning the range $[\ain,\aout]$ (the actual number of field point is unimportant). We see that the accuracy on the potential reaches the computer precision for less than a thousand source point with the GM-rule, and for only a hundred spectral coefficients for the GCL-rule. Note that the relative precision is better than $0.1 \%$ with the CT-rule for a dozen source points only, which is remarkable. Also, we observe that the GCL method does not exhibit a spectral convergence, for reasons mentioned above but appears very efficient. Indeed, these performances are noticeably reduced if quadratures are performed in the $a$-space.

\section{Comparison with the classical FFT-methods}

It is interesting to compare the accuracy of the density splitting method with that of the classical FFT method (e.g Binney \& Tremaine 1987) based on $N \times N$ polar cells. Results obtained on the same test-model are reported on Fig. \ref{fig:errors} as open and filled circles (see below). We see that the precision is rather poor even for large $N$, and is definitively lower than that of the splitting method with the CT-rule.

\begin{figure}[h]
\epsscale{0.45}
\plotone{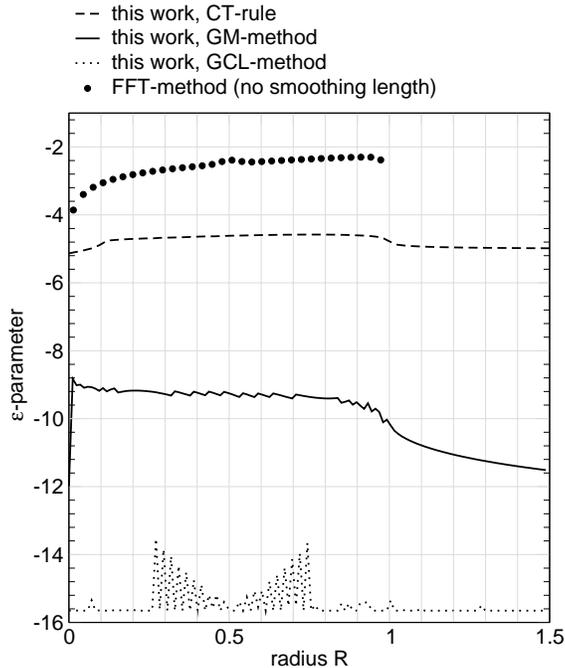}
\caption{$\epsilon$-parameter versus the radius $R$ with the splitting methods ({\it lines}) and with the FFT-method ({\it filled circles}), for $N=128$. For the FFT-method, the disc inner edge is set to $10^{-2}$.}
\label{fig:errorsFFT}
\end{figure}

Let us remind that the FFT-method is probably the most widely method used in simulations of self-gravitating discs to compute the gravitational potential (Binney \& Tremaine 1987). The only advantage of the FFT-method seems to be its great rapidity, $N \log(N)$ order in time. Low computing time is a fundamental requirement if the Poisson equation is to be coupled with other equations, as it is the case in general. One must however realize that the FFT-method has a few major drawbacks. First, it is first-order accurate, as can be seen in Fig. \ref{fig:errors},  that is, one order less than the splitting method with the CT-rule. This means that a huge amount of source/grid points is required before reaching great accuracy. If we extrapolate data shown in the plot (and ignoring loss of significance and round-off errors that would impose a saturation of $\langle \epsilon \rangle$ well above $\epsilon_{\rm CP}$), we find that the FFT-method would need $N \simeq 10^{14}$ to reach the computer precision (that is, $10^{28}$ polar cells). Second, it is a particle-type method (e.g. Hockney \& Eastwood 1988): each cell is made homogeneous, and converted into a particle with arbitrary assignment of both location and mass density. The precision of the FFT-method is apparent and fortunate (assignment errors almost cancel or compensate). Third, the use of the FFT-method in cylindrical coordinates requires a logarithmic spacing of grid points. This means that i) the origin can not be included, ii) the outer disc has always a much lower resolution than the inner disc (by a factor $N$), and iii) many interpolations are necessary when other equations are solved on a different grid (one has to pass from one grid to the other). It is true that matter in astrophysical discs is generally concentrated at the inner edge, but disc self-gravity concerns regions rather located near the outer edge. Fig. \ref{fig:errorsFFT} shows $\epsilon$ versus $R$ for $N=128$ in the example considered before. We see that the accuracy is better in the inner disc than in the outer disc with the FFT-method, whereas it is uniform with the density splitting method (whatever the quadrature rule).

Fourth, the local contribution to the gravitational potential (that is the contribution of the cell containing the field point $R$) is treated in an artificial way. In this purpose, a smoothing length $\lambda$ is often introduced, and its value is chosen arbitrarily. Results obtained for $\lambda=0.01$ are shown in Fig. \ref{fig:errors} (open circles). Hence, potential values are smoothing length-dependent. A slightly better accuracy is obtained with the prescription by Binney \& Tremaine (1987) (filled circles in Fig. \ref{fig:errors}), where the local contribution is estimated analytically, without any $\lambda$. However, as some authors have noticed (e.g. Caunt \& Tagger 2001), the FFT-method can trigger numerical instabilities in hydrodynamic codes, probably because the order of this method is lower (and errors larger) than the order (generally two) of difference schemes used in other fluid equations.

\section{Concluding remarks}

In this paper, we have discussed the possible implementation and performances of the splitting method for razor-thin, axially symmetric discs. In particular, we have emphasized the role of a space mapping in order to rise the accuracy (or effective order) of advanced quadrature rules. We have noticed the important weaknesses of the classical FFT-methods, especially in terms of accuracy, by direct comparisons. Obviously, our method is characterized by much longer computing times. Note however that a few grid points are necessary to reach a high precision. Besides, the method is easily parallelizable.

Several extensions and improvements to the present method could be brought. For instance, other space mapping than proposed can probably work well. However, it is hard to find a sampling function that makes the integrands and derivatives finite (and possibly zero) at the field point without increasing its wings, which is numerically uncomfortable (for instance, this can easily be shown with a $w$-function of the form $|a-R|^{1/n}$ for large $n$). Besides, depending on the global surface density profile in the disc, it can be more appropriate to replace the homogeneous contribution (see Sect. 2) by an another analytical contribution (for instance $\Sigma \propto 1/a$ instead of a constant as done here, provided kernels can be integrated). Finally, the method can easily be extended to treat non-axially symmetrical systems (Pierens \& Hur\'e 2005), and even tri-dimensional mass distributions. We plan to open soon to the scientific community an Internet site devoted to this question.

The splitting method should be useful in many problems where great accuracy is needed and/or fine physical effects must be investigated. For instance, we believe that significant progresses could be made in numerical simulations of planetary migrations which do show a great sensitivity to the field and potential values, even for low mass discs (e.g. Nelson \& Benz 2003).

\section*{Acknowlegments}

We thank the referee for valuable comments.

\appendix

\section{Splitting method for the potential}
\label{sec:appenA}

According to notation, the general expression for the potential in the equatorial plane of a
disc with inner edge $\ain$ and $\aout$ is (e.g Durand, 1964):
\begin{equation}
\Psi=-2G\int_{\ain}^{\aout}{\sqrt\frac{a}{R}k\Sigma(a) {\bf K}(k)da},
\label{eq:potential}
\end{equation}

For a constant surface density $\Sigma_0$, Eq.(\ref{eq:potential}) reads
\begin{equation}
\Psi^{\rm homo.}(R)=-2G\Sigma_0\int_{\ain}^{\aout}{\sqrt{\frac{a}{R}}k \elik(k)da}
\label{eq:potential2}
\end{equation}
For $\ain \le R \le \aout$, we separate this integral into two integrals, leftward and rightward to the fieldpoint where $k=1$, namely
\begin{equation}
\Psi^{\rm homo.}=-2G\Sigma_0\left[\int_{\ain}^R\sqrt{\frac{a}{R}}k \elik(k)da+\int_R^{\aout}\sqrt{\frac{a}{R}}k \elik(k)da\right]
\end{equation}
We now set $u=\frac{a}{R} \le 1$ in the first integral and $v=\frac{R}{a} \le 1$ in the second one. With the Gauss transformation (Gradshteyn \& Ryzbik 1980)
\begin{equation}
{\bf K}\left(\frac{2\sqrt{u}}{1+u}\right)=(1+u) {\bf K}(u)\qquad u<1,
\end{equation}
Eq.(\ref{eq:potential2}) becomes
\begin{equation}
\Psi^{\rm homo.}(R)=-4G\Sigma_0 R\left(\int_{\ain/R}^{1}u{\bf
K}(u)du-\int_{1}^{R/\aout}\frac{{\bf K}(v)}{v^2}dv\right),
\end{equation}
and so
\begin{equation}
\Psi^{\rm homo.}(R)=-4G\Sigma_0 R\left[\frac{\aout}{R}\elie{\left(\frac{R}{\aout}\right)}-\elie{\left(\frac{\ain}{R}\right)}+\left(1-\frac{\ain^2}{R^2}\right)\elik{\left(\frac{\ain}{R}\right)}\right]
\end{equation}

\section{Successive derivatives}
\label{sec:appenB}

From Eq.(\ref{eq:w}), we have
\begin{equation}
\frac{da}{dw} = 2 \sqrt{\aout |R-a|}=
\begin{cases}
-2w\aout > 0 & \text{if} \quad R > a,\\
0 & \text{if} \quad R=a, \\
2w\aout >0 & \text{if} \quad R < a.
\end{cases}
\end{equation}
 
Let $\kappa$ be either $\kappa_\Psi$ or $\kappa_g$, we have
\begin{flalign}
\nonumber
\frac{d}{dw} \left( \delta \Sigma  \times  \kappa  \times |w| \right) & = |w|  \times \frac{d}{dw} \left( \delta \Sigma  \times  \kappa \right) \pm \delta \Sigma  \times  \kappa \\
\nonumber
& = |w| \times \frac{d}{da} \left( \delta \Sigma  \times  \kappa \right) \times \frac{da}{dw} \pm \delta \Sigma  \times  \kappa \\
& = 2 w^2 \aout \times \frac{d}{da} \left( \delta \Sigma  \times  \kappa \right) \pm \delta \Sigma  \times  \kappa
\end{flalign}
for the first derivative,
\begin{flalign}
\nonumber
\frac{d^2}{dw^2} \left( \delta \Sigma  \times  \kappa  \times |w| \right) & = \frac{d}{dw} \left[ 2 w^2 \aout \times \frac{d}{da} \left( \delta \Sigma  \times  \kappa \right) \pm \delta \Sigma  \times  \kappa \right]\\
\nonumber
&= 4 w \aout \frac{d}{da} \left( \delta \Sigma  \times  \kappa \right) + 4 |w|^3 \aout^2 \frac{d^2}{da^2} \left( \delta \Sigma  \times  \kappa \right) \pm 2 |w| \aout \frac{d}{da} \left( \delta \Sigma  \times  \kappa \right)\\
&= 2 \left( 2 w \pm |w| \right) \aout \frac{d}{da} \left( \delta \Sigma  \times  \kappa \right) + 4 |w|^3 \aout^2 \frac{d^2}{da^2} \left( \delta \Sigma  \times  \kappa \right)
\end{flalign}
for the second derivative is, and
\begin{flalign}
\nonumber
\frac{d^3}{dw^3} \left( \delta \Sigma  \times  \kappa  \times |w| \right) & = \frac{d}{dw} \left[
 2 \left( 2 w \pm |w| \right) \aout \frac{d}{da} \left( \delta \Sigma  \times  \kappa \right) + 4 |w|^3 \aout^2 \frac{d^2}{da^2} \left( \delta \Sigma  \times  \kappa \right) \right]\\
\nonumber
&=  6 \aout \frac{d}{da} \left( \delta \Sigma  \times  \kappa \right) + 8 w \aout^2 (|w| \pm 2w) \frac{d^2}{da^2} \left( \delta \Sigma  \times  \kappa \right) \\
& \qquad  \qquad \qquad+ 8 w^4 \aout^3  \frac{d^3}{da^3} \left( \delta \Sigma  \times  \kappa \right)
\end{flalign}
for the third derivative.
\end{document}